\documentclass[
    ,final            
  ,numberedheadings 
  ]
  {aipproc}

\layoutstyle{6x9}

\begin{document}

\title{Accretion and Ejection---The GR/MHD View}

\classification{95.30.Qd, 95.30.Sf, 96.50.Tf, 97.60.Lf, 98.58.Fd}
\keywords      {}

\author{Julian H. Krolik}{address={Department of Physics and Astronomy,
        Johns Hopkins University, Baltimore MD 21218 USA}}

\author{John F. Hawley}{address={Department of Astronomy, University of
        Virginia, Charlottesville VA 22904 USA}}

\begin{abstract}

     The inward flow of matter through accretion disks is driven by MHD
turbulence.  Global general relativistic MHD simulations shed quantitative
light on this process, revealing a number of aspects of accretion previously
unrecognized.  Among them are strong stresses in the marginally stable and
plunging regions near the black hole and electromagnetically-dominated
conical relativistic jets that can form spontaneously from the accretion
flow.  The energy release associated with both of these effects can
significantly augment the classical energy-release estimates based on
purely hydrodynamic models.

\end{abstract}

\maketitle


\section{Old Points of View and New}

    Speaking in Sicily, home of many distinguished relics of the
Classical world, it would only be appropriate to begin this talk with a
reference to the classical view of accretion disks around black holes
and the outflows they drive.  In this view, created in three seminal
papers published thirty-plus years ago, matter drifts slowly inward
through accretion disks due to the action of a mysterious ``viscosity"
while dissipating an amount of energy per unit mass that is a function
solely of the black hole's spin.  More quantitatively, the ``viscous"
stress is assumed, on the basis of dimensional analysis, to be $\alpha p$,
where $\alpha$ is some unknown number less than or comparable to unity,
and $p$ is the internal pressure of the accretion flow \cite{SS73}.
The reason why the heat liberated is supposed to be a function of black
hole spin alone is that all forces are imagined to end at the innermost
stable circular orbit (the ISCO), so that the radiative efficiency
can be identified with the binding energy per unit mass of that orbit
\cite{NT73}.  Meanwhile, relativistic jets are driven outward along the
rotation axes of spinning black holes, even in the absence of ongoing accretion,
as frame-dragging creates an effective electric field parallel to the
large-scale magnetic field threading the hole's event horizon \cite{BZ77}.
The luminosity of these jets is controlled entirely by the strength of
the field on the horizon and the spin of the black hole.

     In recent years, this classical view has been substantially revised.
The stress that transports angular momentum outward so that matter can
move inward is now thought to be the result of fluctuating, but correlated
magnetic fields: orbital shear forces $\langle B_r B_\phi\rangle <
0$ whenever the orbital frequency gradient $d\Omega/dr < 0$.
Because this stress has nothing to do with the sort of kinetic processes
that the name ``viscosity" connotes, unlike a conventional fluid
viscosity, it does not necessarily dissipate any energy where the stress
takes place.  That the field is strong enough to bring matter inward at
a reasonable rate is due to continual stirring by a robust underlying
MHD instability, the magneto-rotational instability \cite{BH91,BH98}.
Because this instability has a very rapid linear growth rate (comparable
to the orbital frequency) and its conditions for growth are very modest
(good enough electrical conductivity to justify the MHD approximation, a
seed magnetic field with energy density no more than the total pressure),
it should be present throughout any disk surrounding a black hole,
as well as in the great majority of disks around other, less extreme,
gravitating objects.  Thus, electromagnetic stress is the primary
vehicle for angular momentum transport in accretion disks.

       A corollary of attributing disk stresses to magnetic fields is
that the classical boundary condition at the ISCO is likely inappropriate.
Its heuristic basis rested on purely hydrodynamic reasoning: as
matter accelerates inward through the ISCO, passing from a domain of
stable orbits to one where they are unstable, its density falls and its
pressure drops.   It would be hard to envision circumstances in which the
small amount of matter resident inside the ISCO could exert much force
upon the far heavier disk outside.   However, as first recognized long
ago \cite{T74} and recently recalled to mind \cite{K99,G99}, magnetic
fields severely undercut this argument---their effectiveness at exerting
stresses has nothing to do with the inertia or pressure of the matter.
Consequently, there is no particular reason for their stresses to cease
at the ISCO.  In fact, if one estimates the magnetic stress at the ISCO
by supposing that the field is strong enough to drive accretion at the
requisite rate in the main disk body and is then advected inward by
flux-freezing, the magnetic stress there is likely to be quite sizable
\cite{K99,G99}.

       A second corollary of this picture is that magnetic fields arrive at
the event horizon because they are brought there by the accretion
flow.  The field strength at the horizon, the basic parameter of the
Blandford-Znajek \cite{BZ77} mechanism, should then be determined by
the history and current state of accretion.

\section{Numerical Simulations}

      Analytic tools for studying turbulence are very limited; large-scale
numerical simulation is the only technique we have for obtaining
quantitative results.  In the past few years both the capabilities
of the hardware and the quality of the computer codes available for
this effort have improved dramatically, leading to rapid improvements
in accretion disk studies.  Relatively recently, it was possible to
study disks only in the shearing-box approximation, in which an annular
segment is modelled by straightening it out into a rectangular geometry,
while retaining a plane-parallel shear, a tidal potential and a Coriolis
force to approximate the dynamics of orbital motion.  Given the nature
of this approximation, it made sense to perform these simulations in
the Newtonian limit.

       Today's codes, on the other hand, can treat global disks, initially
with Newtonian gravity and now in full general relativity
\cite{GMT03,DH03,A06,M06}.  Under the assumption of ideal MHD (modulo the
inevitable dissipation that occurs on the gridscale), they can follow
fully three-dimensional dynamics for very long timescales: durations
$\sim 10^4 M$ (for black hole mass $M$, setting $G=c=1$).  These codes
do an excellent job of conserving angular momentum, whether associated
with the matter or electromagnetic fields, but their fidelity to Nature
is much more limited in regard to thermodynamics and radiation.  The
De~Villiers-Hawley \cite{DH03} and Anninos et al. \cite{A06} codes evolve an
internal energy equation and treat the gas equation of state as adiabatic
with $\gamma = 5/3$, adding entropy only in association with shocks
through the use of an artificial bulk viscosity.  They therefore
capture neither the bulk of the heat production that is a concomitant of
accretion nor the photon radiation that cools the gas and produces the
light we see.  In a sense, the first omission compensates for the second.
The Gammie-McKinney-T\'oth \cite{GMT03} and Mizuno et al. \cite{M06}
codes evolve a total energy equation.  By construction, losses in kinetic
and magnetic energy are transferred into heat, thereby modeling the
turbulent dissipation process, but like the others, these codes also
wholly ignore radiation.

      Most of the global disk simulations done to date by various
groups assume either an initial large scale vertical net field threading
the disk (most of these have been Newtonian simulations), or an initial
disk that has no net flux, i.e., all field lines closing within the disk.
Simulations with strong net flux generally have little difficulty producing
jets and outflows, but we have no way of knowing whether such a strong
field passing through the central region of the accretion flow is
realistic.  The zero net
flux initial condition is the simplest in terms of guessed parameters,
as it means the field is simply zero all around the outer boundary.
This is the approach we have adopted for our simulations.

      To avoid the numerical difficulties arising from subsonic flow
across the boundary, all our simulations place their entire mass supply
within the computational volume.  This fact has the consequence that
angular momentum removed from the inner accretion flow is given to the
mass in the outer disk, which must then move outward.  In other words,
only the part of the flow inside a certain radius (variously 12--$25M$,
depending on the simulation) actually participates in a genuine accretion
flow.  The outer boundary is generally located much farther out ($r_{\rm
out} \simeq 100 M$).

\section{Properties of the Accretion Flow}

\begin{figure}\label{fig:stress}
  \centerline{\includegraphics[height=.3\textheight,angle=90]{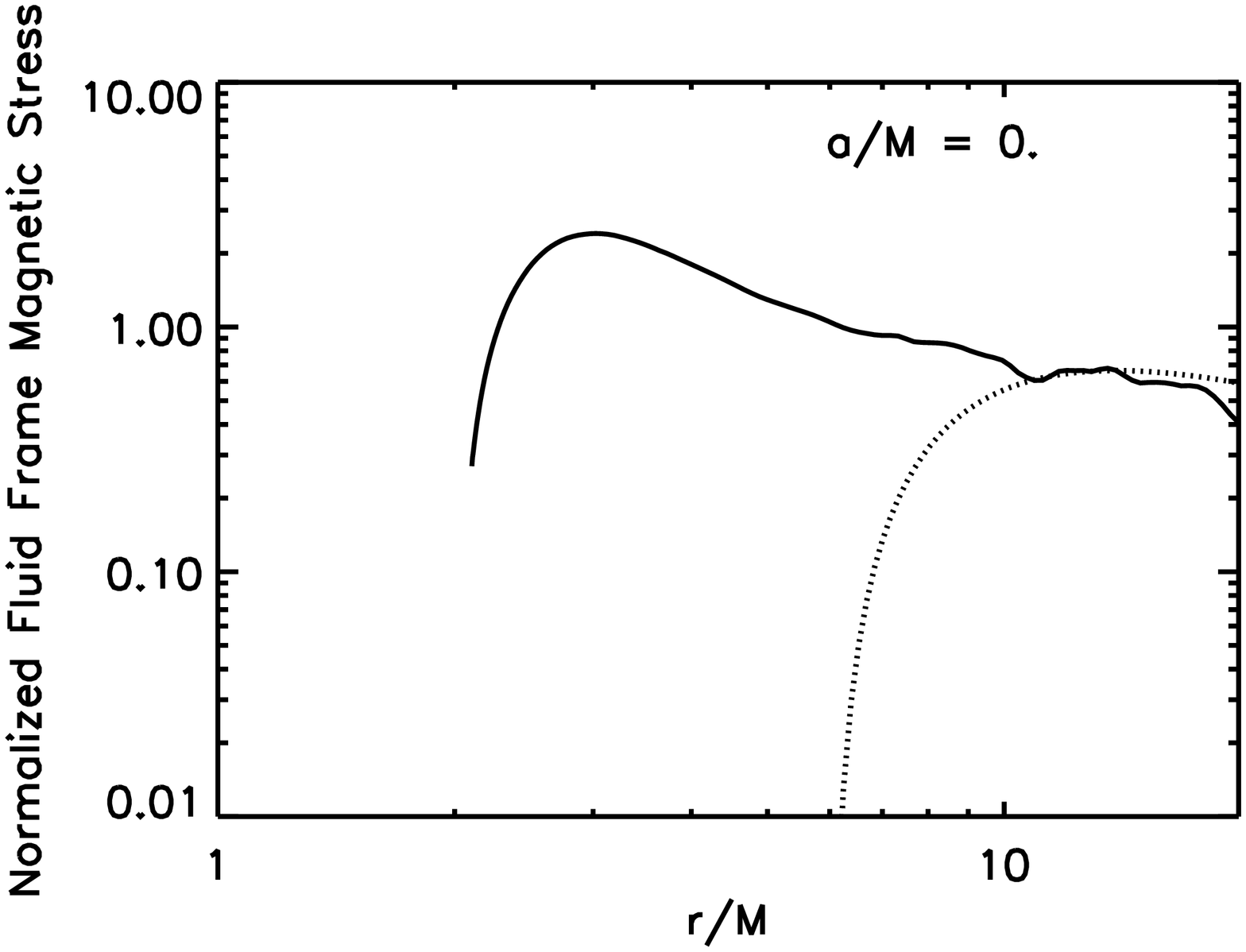}
              \includegraphics[height=.3\textheight,angle=90]{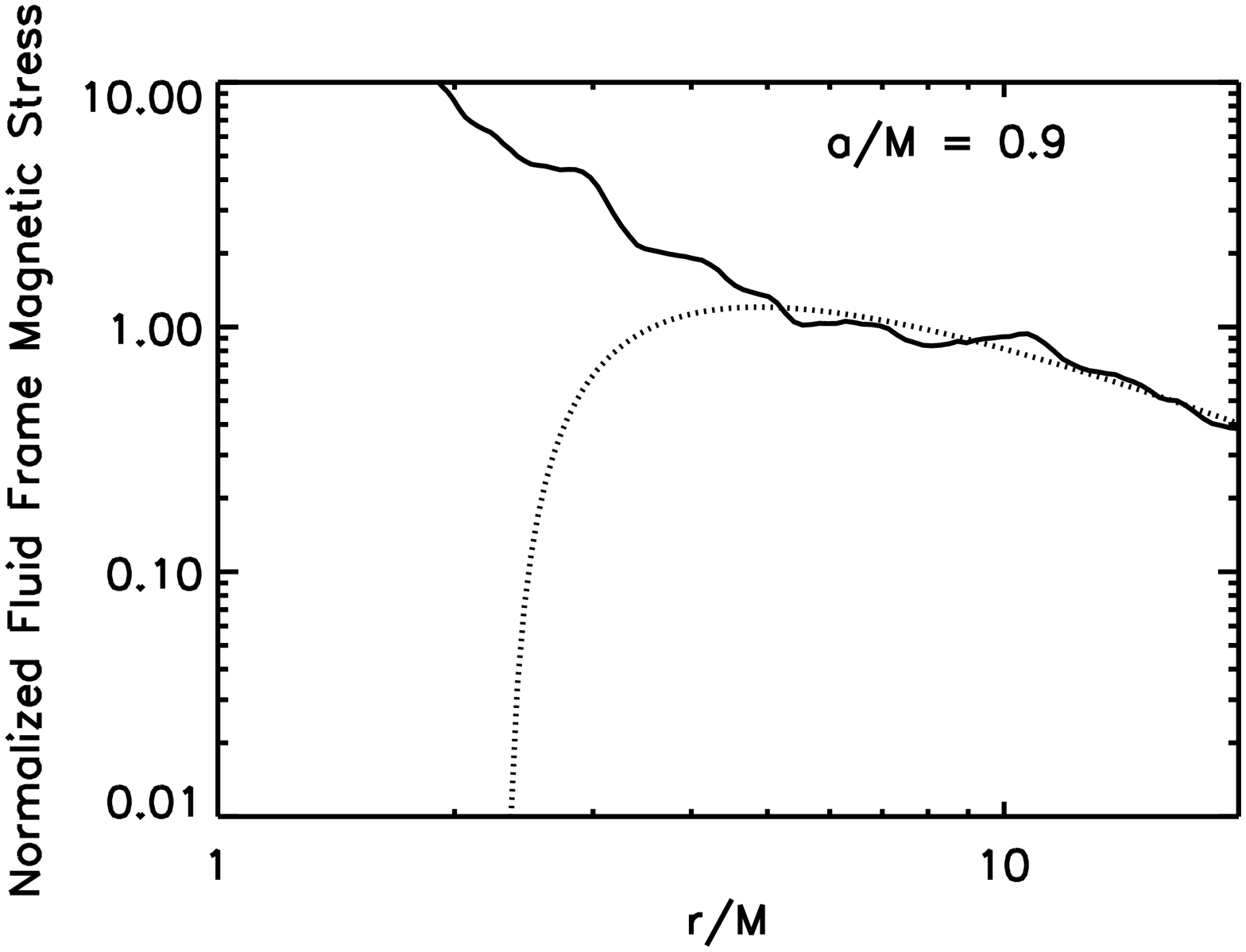}}
  \caption{Fluid-frame stress, integrated over spherical
  shells.  Left panel shows a simulation with a non-rotating black hole,
  right panel one with a black hole having a spin parameter $a/M = 0.9$.
  Solid curves are the electromagnetic stress as found in the simulations at
  a particular time; dotted curves are the prediction of the Novikov-Thorne
  model for an accretion rate matching the time-average of that simulation.}
\end{figure}

    A number of results of qualitative importance have been derived
from these studies.  Most notably, the order-of-magnitude estimates
predicting that magnetic stresses at the ISCO should be important
have been strongly vindicated \cite{K05}.  As Figure~\ref{fig:stress}
demonstrates vividly, whereas the classical Novikov-Thorne picture
predicts that the stress ends abruptly at the ISCO, the actual MHD
stresses continue quite smoothly inward toward the black hole.  In the
case of the non-rotating black hole, there can be no outward stress at the
event horizon because there is no source of angular momentum to supply the
outward flow of angular momentum that constitutes the stress.  When the
black hole rotates, however, the black hole itself is an ample reservoir
of angular momentum, and there is no reason why the stress must end
somewhere outside the event horizon.  Although this process might appear
to violate causality, there is no such violation in reality: This process
can equally well be thought of as an inward flow of negative angular
momentum facilitated by the rotating black hole's ergosphere, where both
matter and electromagnetic fields can have negative energy-at-infinity.

\begin{figure}\label{fig:alpha}
  \centerline{\includegraphics[height=.3\textheight,angle=90]{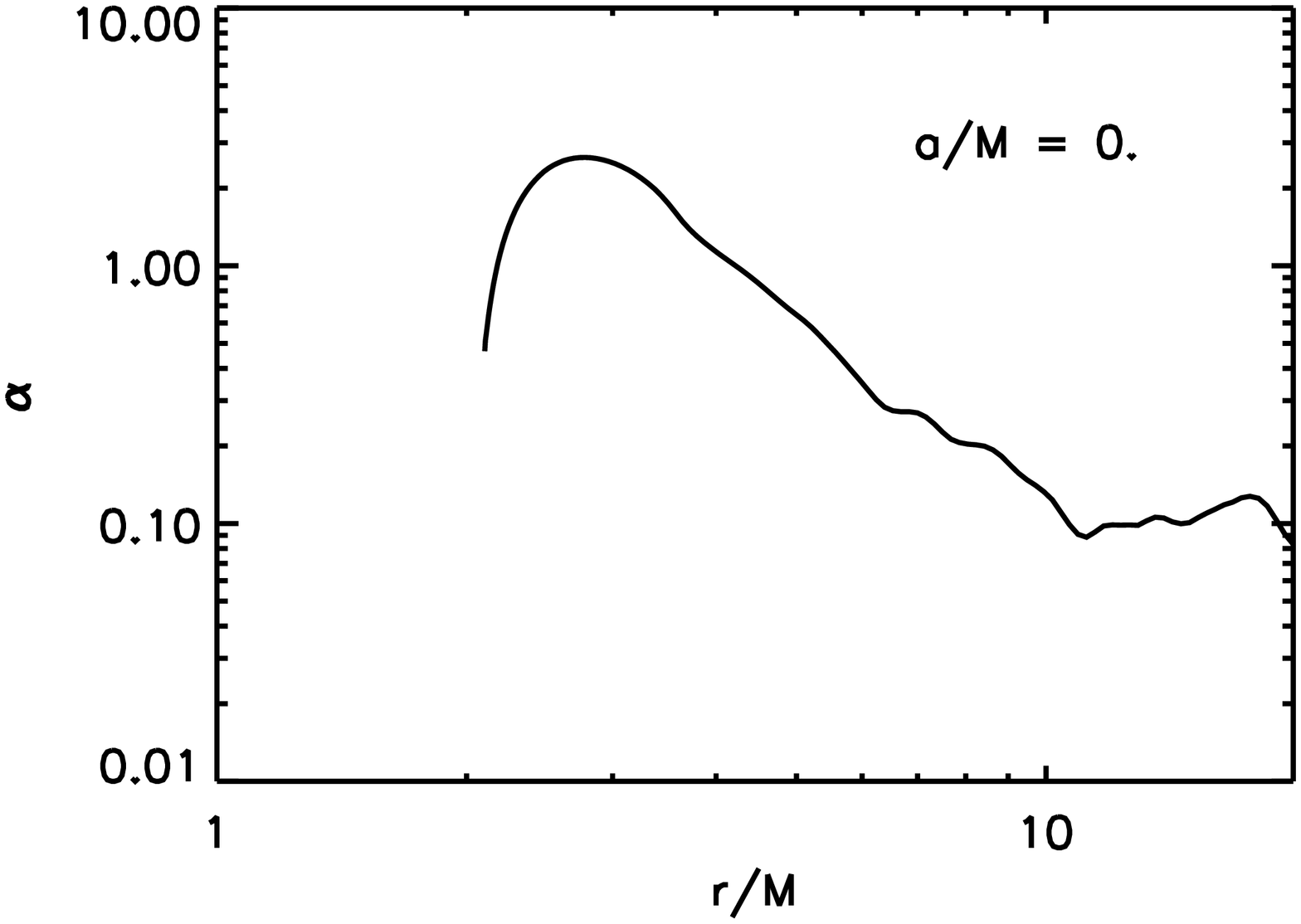}
              \includegraphics[height=.3\textheight,angle=90]{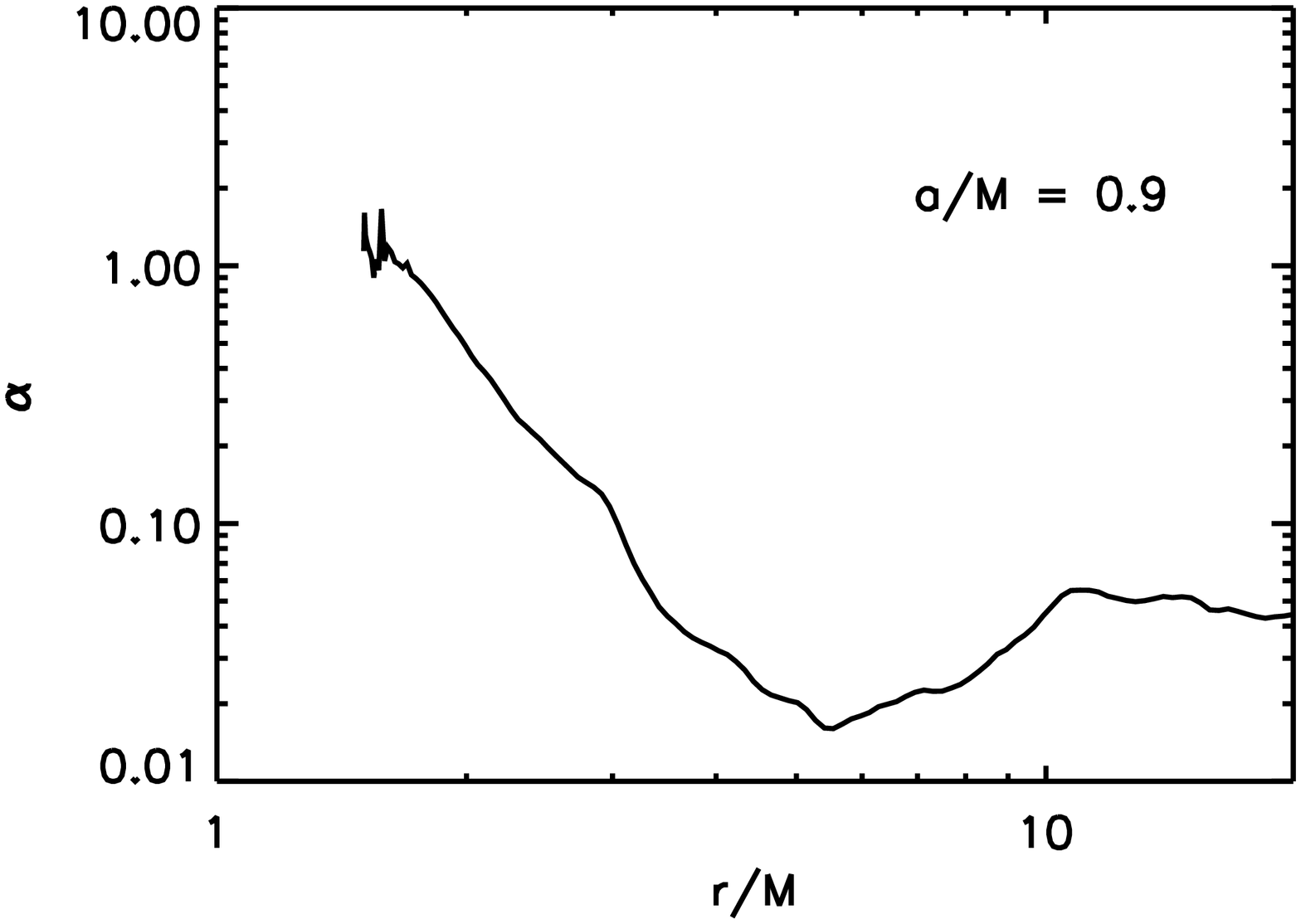}}
  \caption{Ratio of shell-integrated electromagnetic stress to shell-integrated
  pressure, all evaluated in the fluid frame at a particular coordinate time.
  Left panel shows a simulation with a non-rotating black hole, right panel one
  with a black hole having a spin parameter $a/M = 0.9$.  Note that the ISCO
  for $a/M = 0$ is at $6M$, while for $a/M=0.9$, it is at $2.32M$.}
\end{figure}

      Regarding the Shakura-Sunyaev $\alpha$ parameter as merely a measure
of the stress in pressure units, we find that it rises rapidly with diminishing
radius, beginning well outside the ISCO and continuing through the plunging
region (Fig.~\ref{fig:alpha}).  In the disk body, $\alpha \simeq 0.02$--0.1.
As one moves closer to the ISCO, the pressure diminishes (in
relative terms) even while the magnetic stress increases.  The result
is that the nominal $\alpha$ is 5--10 times larger near the ISCO
and another factor of 5 or more larger still deep inside the plunging
region.

     Throughout the disk, wherever there are stresses transporting angular
momentum radially, there is also work done by one ring on its neighbor.
The energy flow into a ring due to the imbalance between the work it does
and the work done on it is not necessarily cancelled by the net change
in its orbital (and electromagnetic energy) content due to accretion.
When Novikov and Thorne first calculated the radial profile of dissipation
on the basis of energy and angular momentum conservation alone, it was
exactly this mismatch that lay at the basis of their work.  Stress at
the ISCO therefore alters the total dissipation possible in the disk
outside that radius \cite{AK00}.

     Recently, we have attempted to estimate the total
radiative efficiency, including heat liberated inside the ISCO (and
therefore beyond the reach of the Agol-Krolik \cite{AK00} formalism)
and also allowing for losses due to photons following orbits that bring
them to the black hole rather than infinity \cite{BHK06}.  Unfortunately,
as mentioned earlier, we do not actually compute the dissipation rate
directly.  Instead, we must seek some way of estimating the dissipation
rate based on local physical quantities.  Although we cannot prove
that a diagnostic based on guesswork is correct, we can eliminate
those that must be wrong: in the disk body, where the inflow time is
long compared to the thermal time, energy conservation in the context
of a time-steady disk demands a definite dissipation profile.  With no
stress at the ISCO, this is the profile predicted by Novikov and Thorne;
with stress, this form is generalized, as worked out by Agol and Krolik.
Any candidate prescription for estimating the dissipation must therefore
match the profile required by energy conservation in the time-steady
portion of the disk.   After trying a number of possibilities, all
heuristically linked to energy production one way or another, we found
that the best agreement with the requirements of energy conservation
in a time-steady disk was achieved with the assumption that there
is a uniform resistivity
$\eta$, so that the rate of heat generation is $\eta J^\mu J_\mu$, where
$J^\mu$ is the four-current density.  To compute the luminosity observed
at infinity, we then assumed that photons are
emitted isotropically in the fluid frame at a rate proportional to that
dissipation function and traced their trajectories until they reach
infinity or are swallowed by the hole.

     As it turns out, most of the dissipation predicted by this rule
takes place outside the ISCO.  In addition, many of the photons radiated
from within the ISCO are on orbits that ultimately lead to capture by the
black hole.  Consequently, the bulk of the additional radiative efficiency
seen by distant observers has its origin near and just outside the ISCO.

     Table~\ref{tab:diskeff} shows the numbers derived from a single
snapshot from each of four simulations.  The entries in the second
column are the efficiencies predicted by the Novikov-Thorne model,
but after allowance for photons captured by the black hole.  They are
therefore smaller than the standard quoted values, by a very small
amount for $a/M = 0$, where the ISCO is relatively far from the horizon,
by a more substantial amount (about a $10\%$ fractional reduction) for
the case of $a/M = 0.998$.  Ranges rather than single values are shown
because some photon orbits intersect the equatorial plane before heading
off to infinity.  When they do, they may interact with the matter in
the accretion flow.  Scattering or absorption followed by reradiation
puts the photons onto a new orbit that may or may not go to infinity.
Which end-result obtains depends very much on the details.  To allow for
this very model-dependent uncertainty, we quote ranges in which the lower
number is the result of assuming that none of the energy from photons
on plane-crossing orbits reaches infinity while the upper number is the
result of the opposite assumption.

\begin{table}
\begin{tabular}{lrr}
\hline
 \tablehead{1}{c}{b}{$a/M$} &
 \tablehead{1}{c}{b}{$\epsilon_{NT}$} &
 \tablehead{1}{c}{b}{$\epsilon_{MHD}$}\\
\hline
0.0   & 0.055--0.056 & 0.067--0.07 \\
0.5   & 0.077--0.079 & 0.13--0.14  \\
0.9   & 0.137--0.145 & 0.16--0.18  \\
0.998 & 0.250--0.290 & 0.29--0.41  \\
\hline
\end{tabular}
\caption{Estimated radiative efficiency from accretion, in rest-mass units.}
\label{tab:diskeff}
\end{table}

     Comparing the Novikov-Thorne numbers with those based on this estimate
from the simulations indicates that interesting augmentations in radiative
efficiency may be achieved.  The greatest fractional increase is for
$a/M \simeq 0.5$ because increasing spin results in a trade-off: more
stresses at small radii, but also larger probabilities of photon capture.

\section{Outflows}

      One of the most interesting results from recent disk simulations
has been the discovery that Poynting-flux jets can form spontaneously
within a cone aligned with the black hole's rotation axis.
When accreting matter approaches the black hole's event horizon,
there is an effective resistivity due to general relativistic kinematics
that lets matter cross field lines and enter the black hole, even while
the field lines remain just outside.  One way to understand this process
is to recall that gravity immediately outside the event horizon is so
powerful that it can overcome the Lorentz force binding electric charges
to their Larmor orbits, forcing them to deviate away from their usual
guiding center motion along a field line.

       Field lines thus shorn of their inertia find it easy to float upward.
In fact, if there were no magnetic field at high latitudes around the event
horizon, the $\nabla B^2$ force would quickly cause them to spread into that
region.  Once there, they can expand upward and outward.  Indeed, we find
that in simulations with non-rotating black holes a small amount of accretion
can suffice to create a magnetic field in a cone around the rotation axis
that is almost exactly radial and stretches to the outer boundary of the
simulation.

       Very little matter penetrates into this cone around the rotation
axis for the simple reason that there is a strong centrifugal barrier.
Even matter with the small amount of angular momentum permitting an orbit
at the ISCO has too much angular momentum to approach the axis above
the black hole.   Numerical difficulties prevent us from quantifying
just how small the matter density is, but we can say with confidence
that it must be very small.  As a result, this region is strongly
magnetically-dominated and highly relativistic, moving outward with a
Lorentz factor that we can only bound from below at $\simeq 5$--10.
In other words, we have found that, in the right circumstances,
relativistic jets can spontaneously form as a result of accretion.

\begin{figure}\label{fig:fieldlinerot}
  \centerline{\includegraphics[height=.3\textheight,angle=90]{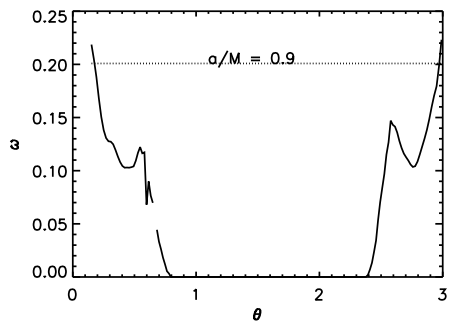}}
  \caption{Radially-averaged field line rotation rate at a particular time
  as a function of polar angle in the jet when $a/M = 0.9$ (solid curve).
  The dotted line shows the rotation rate predicted in the simplest
  version of the Blandford-Znajek model.}
\end{figure}

        When the black hole rotates, the frame-dragging at small radius
pulls the field into rotation (see Fig.~\ref{fig:fieldlinerot}) and
creates a transverse component.  Moving transverse magnetic field,
of course, carries Poynting flux.  In its fundamental character, this
mechanism resembles the classical Blandford-Znajek mechanism.   However,
this version differs in that, rather than operating independent of
accretion, the very existence of the field is due to accretion, and its
strength is regulated by it.  In addition, we are unrestricted in black
hole spin rate (the original model calculation involved a perturbation
expansion in $a/M$) and compute the field shape self-consistently.

       The magnitude of the electromagnetic luminosity associated with this
jet can be sizable.   Phrased in terms of efficiency in rest-mass units,
when the black hole spins rapidly the jet power can be competitive
with the radiative luminosity expected from the accretion disk
(Table~\ref{tab:jeteff}, taken from \cite{HK06}; see also \cite{MG04}).

\begin{table}
\begin{tabular}{lrr}
\hline
 \tablehead{1}{c}{b}{$a/M$} &
 \tablehead{1}{c}{b}{$\epsilon_{NT}$} &
 \tablehead{1}{c}{b}{$\epsilon_{jet}$}\\
\hline
-0.9  & 0.039  & 0.023 \\
0.0   & 0.057  & 0.0003 \\
0.5   & 0.081  & 0.0063 \\
0.9   & 0.16   & 0.046  \\
0.95  & 0.19   & 0.072  \\
0.99  & 0.26   & 0.21 \\
\hline
\end{tabular}
\caption{Poynting flux jet efficiency, in rest-mass units.  The Novikov-Thorne
efficiency shown for comparison is the conventional one, identical to the
specific binding energy at the ISCO.}
\label{tab:jeteff}
\end{table}

     Not all magnetic geometries are this efficient, however.
The numbers in Table~\ref{tab:jeteff} were computed on the basis
of an initial magnetic field shape that is ``dipolar" in the sense that
the field lines wrap concentrically around the pressure maximum of the
original matter distribution.  When the initial field is, instead, ``quadrupolar"
in the sense that the field is divided into two sectors, with one set
of field loops centered on a ring offset above the equatorial plane, the
other positioned symmetrically below, reconnection in the inner accretion
flow is so efficient that the Poynting flux in the jet is greatly reduced.

\section{Summary}

    With the progress of the past fifteen years, there have been
substantial advances in our understanding of the physics of accretion disks.
The origin of the stress that plays such a fundamental role in driving
accretion is no longer unknown---it is MHD turbulence, driven by a
powerful and robust instability.

    Although the equations governing such a system are strongly nonlinear,
they are based on very well-understood physics: nothing more than
classical electromagnetism and relativistic gravity.  Because we can
build upon such a solid foundation, it makes sense to undertake the
arduous task of constructing large-scale simulations.  These are now so
feasible that there are several competing groups performing them.

    Much remains to be explored in this rich territory, but already
several important results have been established:

\begin{itemize}

\item Contrary to expectations based on hydrodynamic intuition, MHD
stresses continue to be strong all the way through the ISCO and (if
the black hole rotates) in to the event horizon.  Precise
estimates of the additional efficiency due to these stresses are
not yet available, but we can begin to make estimates.

\item When the magnetic field has the right geometry, relativistic
electromagnetically-dominated jets can form along the black hole
rotation axis.  Driven by the rotation of the black hole, these
jets can carry Poynting flux to infinity at a rate that may increase
the power output per unit rest mass accreted by factors of order
unity relative to that radiated directly from the accretion disk proper.

\end{itemize}


\begin{theacknowledgments}
  We are grateful to all our collaborators on this subject:
Jean-Pierre De Villiers, Shigenobu Hirose, Kris Beckwith, and
Jeremy Schnittman.  This
work was partially supported by US National Science Foundation Grant
AST-0507455.
\end{theacknowledgments}


\end{document}